\documentclass[prb,twocolumn,showpacs,amsmath,amssymb,superscriptaddress,unsortedaddress,floatfix]{revtex4}

\usepackage{graphicx,amsfonts,amsmath}% Include figure files
\usepackage{dcolumn}% Align table columns on decimal point
\usepackage{bm}% bold math
\usepackage{color}
\newcommand\ket[1]{|#1 \rangle}

\newcommand\dmatrix[3]{\langle #1|#2|#3\rangle}
\newcommand\ketbra[2]{|#1 \rangle\!\langle#2|}

\newcommand\tr[1]{\mathrm{Tr} ~ #1}

\begin{document}

\preprint{APS/PRB}

\title{Phonon influence on the measurement of spin states in
double quantum dots using the quantum point contact}% Force line breaks with \\

\author{{\L}ukasz Marcinowski}
\affiliation{%
Institute of Physics, Wroc{\l}aw University of Technology,
50-370 Wroc{\l}aw, Poland
}%
\author{Katarzyna Roszak}%
\affiliation{%
Institute of Physics, Wroc{\l}aw University of Technology,
50-370 Wroc{\l}aw, Poland
}%
\author{Pawe{\l} Machnikowski}%
\affiliation{%
Institute of Physics, Wroc{\l}aw University of Technology,
50-370 Wroc{\l}aw, Poland
}%
\author{Mateusz Krzy{\.z}osiak}
\affiliation{%
Beijing University of Technology, 100124 Beijing, China,
}%

%\keywords{Suggested 
\begin{abstract}
We study the influence of phonon scattering on the noise characteristics
of a quantum point contact coupled to a two-electron
system in a double quantum dot, as proposed for a singlet-triplet 
measurement scheme in a double-dot system. We point out that at low temperatures 
phonon-induced relaxation to the ground state suppresses transitions
to doubly occupied singlet states which are the source of detectable
current fluctuations in this measurement scheme. 
Thus, for a relatively strong electron-phonon interaction
present in the system, the two configurations display the same noise
characteristics. In this way, coupling
to phonons reduces the 
distinguishability between the singlet and triplet configurations.
Under such conditions, the proposed measurement scheme is no longer valid
even though the times of the measurement-induced decoherence of an 
initial singlet-triplet
superposition and of the localization into the singlet or triplet subspace
remain essentially unchanged.
\end{abstract}

% PACS, the Physics and Astronomy
                             % Classification Scheme.
%\keywords{Suggested 
\maketitle

\section{\label{sec:Intro}Introduction}

One of the most promising proposals for solid-state qubit implementation 
is based on the utilization of the spin states of electrons
confined in quantum dots (QDs). The original idea of coding the qubit
in the two states of a single electron spin\cite{loss98} 
still inspires a lot of interest,
since relatively long spin coherence times have been reported, and the experimental 
techniques for the preparation, manipulation, and readout of such qubits 
are being rapidly developed\cite{nowack07,greilich09,kim11}.
The study of two electron spin states in double QDs (DQDs)
is a natural extension of the problem, which serves to examine two-qubit coherence
and inter-qubit interactions\cite{coish05,johnson05,pfund07,stepanenko12,maune12}.
Furthermore, to facilitate electrical control of 
electron-spin qubits, two-spin encoding has been proposed\cite{levy02,petta05,taylor07},
which involves 
spin-singlet and spin-triplet configurations serving as the $|0\rangle$ and $|1\rangle$
qubit states. This approach proves to be promising as well, as is seen, e.g.,
in the recent demonstration
of entanglement between two singlet-triplet qubits\cite{shulman12}.
 
The quantum point contact\cite{beenakker91} (QPC)
measurement of charge states in a lateral DQD
defined by gate potentials in a two dimensional electron gas 
involves monitoring the current
flowing through the QPC which depends on the occupation of the QDs due
to a Coulomb interaction between the electrons confined in the QD and
electrons traveling through the QPC\cite{barrett06,stace04}.
This measurement scenario is a realization of the so called weak
measurement\cite{nielsen00}, 
where the measured system is only weakly coupled to the measuring device.
Contrary to the projective measurement, this measurement is not
instantaneous, 
as both the localization of the QD states into the measurement basis and 
acquiring the data needed to distinguish between the basis states take time.
Apart from the measurement time, another relevant factor is the
attainable distinguishability of states, since 
even after an infinitely long measurement time it may not be possible to
completely distinguish between
the measurement basis states.
On the other hand, a weak measurement is typically less destructive to the measured
system than an instantaneous projective measurement.
Furthermore, such a measurement is the only option in many involved quantum systems
which are hard to access experimentally. Hence, the QD-QPC
measurement setup is commonly used
experimentally to study QD occupations at very low
temperatures\cite{averin05a,meunier06,rogge05,barthel09,cassidy07,bylander05,elzerman04}. 
As we have previously shown, phonon effects do not
interfere with the \textit{charge}  measurement in any significant
way\cite{marcinowski11}, 
since while they strongly affect the
coherence times of QD states, phonons do not affect the localization times
or the distinguishability between the measurement basis states in this setup. 

The measurement of spin states of electrons confined 
in QDs is much more complicated and typically
involves spin-to-charge conversion prior to a QPC measurement of the
charge\cite{meunier06,cassidy07,elzerman04,petta05}. 
An alternative scheme for the direct measurement of the spin symmetry
(singlet--triplet) of two-electron states  
confined in a DQD was proposed in Ref.~[\onlinecite{barrett06}]. %[18].
Here, the quality of the measurement relies
on QPC current noise being different for the singlet and triplet spin symmetries. 
The disparity of current fluctuations is due to the fact that,
according to Pauli exclusion principle, states with both electrons
localized in the same QD are allowed in the spin-singlet configuration, but not
for spin-triplet case.  
Hence,
the electron charge distribution will fluctuate during the measurement
process due to the  
QD-QPC interaction only if the electrons are in the spin-singlet state,
leading to enhanced QPC current noise for this spin configuration. 

\begin{figure}[tb]
  \includegraphics[scale=0.59]{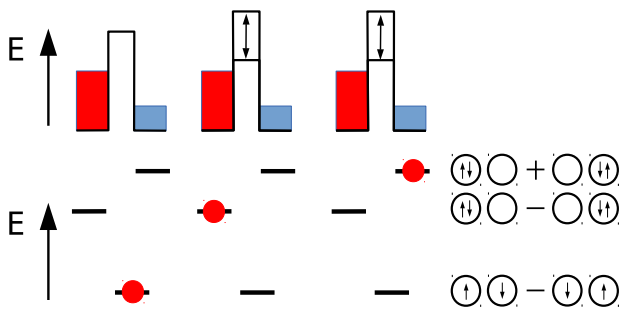}
  \caption{Schematic representation of the 
changes of the QPC barrier height due to Coulomb interaction with different spatial configurations of the three spatially distinct DQD spin singlet states. 
The top part shows the QPC barrier height corresponding to the occupation
of the DQD represented in the bottom part of the figure.
The two doubly occupied states are superpositions of configurations with
charges adjacent or remote from the QPC which results in fluctuations 
of the QPC barrier.}
  \label{fig:measurement}
\end{figure}

In this paper, we study the interplay of phonon-induced effects on
two-electron DQD spin  
states and the QPC measurement of these states in the high bias regime.
In this setup, phonon-assisted
interdot tunneling processes at low temperatures lead to relaxation of electrons in the spin-singlet
configuration to their lowest energy state \cite{grodecka08a}
(with the two electrons located in different dots to minimize the
repulsive Coulomb interaction), which should strongly affect the
distinguishability between 
the singlet and triplet states. 
As we find out, 
while the additional, phonon-induced decoherence channel
obviously increases the rate of dephasing of singlet-triplet
superpositions, it affects neither the time needed for the measured state to localize
in the singlet or triplet state (``collapse'' of the state vector)
nor the data acquisition time needed to perform the measurement. 
We show, however, that 
phonon-assisted transitions counteract the enhanced QPC current
fluctuations in the case of the singlet state by suppressing
the occupation fluctuations. Thus, the carrier-phonon interaction leads to a
reduced distinguishability of two-electron  
spin states. 
When the strength of the phonon-QD interaction is comparable to the
strength of the QPC-QD coupling, the measurement process
is completely suppressed and an extension of the measurement time 
cannot yield any improvement on
the quality of the measurement. 

The paper is organized as follows. In Sec. \ref{sec:Model}, we introduce the system
and define the model to be studied. In Sec. \ref{LindMeq}, we derive the
quantum master equation in Lindblad form for the dynamics of the
DQD-QPC system with 
the electron-phonon interaction included. In Sec. \ref{sec:Method}, we introduce the
stochastic simulation method in the conditional
density matrix formalism,
which allows us to perform simulations of single measurement runs.
The general results are presented in Sec. \ref{sec:Results},
while the noise characteristics are discussed in Sec. \ref{sec:Noise}.
Sec. \ref{sec:Conclusions} concludes the paper.

\section{\label{sec:Model}The system and the Hamiltonian}

We consider two electrons confined in a gate defined lateral
DQD composed of two identical QDs coupled to a QPC, following Ref.~[\onlinecite{barrett06}]. 
The QPC is located near one of the dots (say, right) in such a way
that the current flowing through the QPC is only affected by the
occupation of this one dot.
We assume that the electrons are in the ground state manifold of 
single electron orbital states
and that the excited states are energetically far beyond the double
charging energy. Because of this assumption, only six two-electron
states are taken into account. The four lower energy states involve electrons
confined in separate QDs, one with singlet spin symmetry and three with triplet spin
symmetries. Due to the Coulomb repulsion energy, the other two states which are 
superpositions of doubly occupied QD states and must have singlet spin symmetry
due to the Pauli exclusion principle, are energetically separated from the 
other two-electron states considered.
The electrons tunneling through the QPC interact with the electrons in the right 
QD due to the dependence of the QPC tunneling barrier on
the occupation of the right dot (the larger this occupation, the higher
the tunneling barrier). When the QD electrons are in one of the lower energy 
states, the height of the QPC tunneling barrier is robust and the current flowing
through the QPC is Poissonian (the noise is characteristic for noninteracting
electrons travelling through a time-independent potential barrier). The situation is different
when the DQD is in one of the higher singlet states which are superpositions 
of doubly occupied states, because the height of the tunneling barrier is different
when both electrons are in the right dot and when both electrons are in the left
dot (see Fig.~\ref{fig:measurement}). 
The resulting fluctuations of the barrier height lead to an enhancement of the QPC current 
noise. Since the interaction
between the DQD and the QPC cannot change the electron spin symmetry,
only the low energy singlet state is coupled to higher energy states.
Hence, if the DQD is in the triplet state the QPC current noise remains 
Poissonian throughout the evolution, but if the DQD is in the singlet state,
the QPC induces DQD transitions between different singlet
states, which leads to different noise characteristics (and much greater
current noise).
This difference in the magnitude and the statistical characteristics of the QPC current noise allows 
one to distinguish between the DQD triplet and singlet states, and is the basis
of the measurement scheme proposed in Ref.~[\onlinecite{barrett06}]. 
In order for the non-elastic tunneling events at the QPC to provide sufficient energy
for inducing a transition to a doubly occupied state,
the QPC has to be operated in the high bias regime, that is, the difference between the 
chemical potentials in the source and drain must be larger than the double charging energy.
Apart from this requirement, no fine tuning of the bias voltage is needed, in contrast to the spin-charge 
conversion protocols \cite{reilly07}.

Since QDs are solid state systems,
we include the coupling between the confined electrons
and vibrations of the surrounding crystal lattice (phonons), especially
that the time scales of phonon-related processes in such a system are
comparable to the times over which QPC current traces are 
observed\cite{marcinowski11,cassidy07,bylander05}.
The interaction between the DQD electrons and the phonon environment
cannot change the spin symmetry of the DQD state (similarly as the DQD-QPC 
interaction), hence, it does not couple singlet and triplet states. In fact,
as shown later, phonons induce exactly the same transitions as the DQD-QPC 
interaction and at low temperatures phonon emission from the DQD is expected 
to suppress the QPC induced singlet-singlet transitions. Since the transitions
to higher (singlet) energy states result in the increased QPC current
which serves to distinguish between the singlet and triplet DQD configurations,
such a suppression leads to a diminished distinguishability between the states 
and consequently can undermine the usefulness of the measurement scheme.

The Hamiltonian of the system is given by
%\cite{bib:SiB}
\begin{displaymath}
H_{\mathrm{tot}} = H_{{\mathrm{DQD}}} +
H_{{\mathrm{leads}}} + H_{{\mathrm{tun}}} + H_{{\mathrm{ph}}}
+
H_{{\mathrm{e-ph}}}.
\end{displaymath}
The first term describes two electrons in the DQD structure\cite{barrett06,roszak09}
\begin{displaymath}
H_{{\mathrm{DQD}}} = 
\Delta \sum_{\sigma=\uparrow , \downarrow}
(a_{R\sigma}^\dagger a_{L\sigma} + a_{L\sigma}^\dagger a_{R\sigma}) +
U \sum_{i =R,L} n_{i\downarrow} n_{i\uparrow},
\end{displaymath}
where $\Delta$ is the amplitude of the tunneling between the dots,
 $a_{i\sigma}$, $a_{i\sigma}^\dagger$
are the annihilation and creation operators of an electron in dot $i = \mathrm{R,L}$ with
spin $\sigma = $~$\uparrow,\downarrow$, $n_{i\sigma}=a_{i\sigma}^\dagger a_{i\sigma}$ 
gives the number of electrons with spin $\sigma$ in dot
$i$, and $U$ is the Coulomb charging energy for adding a second electron to a QD.

\begin{figure}[tb]
  \includegraphics[scale=0.7]{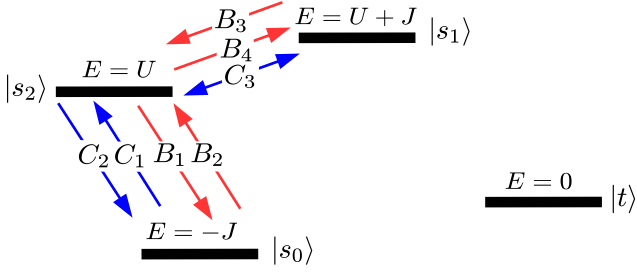}
  \caption{Schematic representation of the energy levels of the system
and the allowed transitions between $H_{{\mathrm{DQD}}}$ eigenstates 
    induced by the interaction of electrons in the DQD with
    the QPC ($C_i$ operators, blue arrows) and with the phonon reservoir 
($B_i$ operators, red arrows).}
  \label{fig:levels}
\end{figure}

The eigenstates of $H_{{\mathrm{DQD}}}$ are 
\begin{subequations}
\label{eqn:states}
\begin{eqnarray}
\label{trypleta}
   \ket{t_0~\!} & = &\frac{1}{\sqrt{2}}
(\ket{\uparrow \downarrow} + \ket{\downarrow \uparrow}),\\
\label{trypletb}
\ket{t_+} &=& \ket{\uparrow \uparrow}, \\ 
\label{trypletc}
\ket{t_-} & = & \ket{\downarrow
  \downarrow}, \\
\ket{s_0} &=&  \xi' \Big(\ket{\uparrow \downarrow} -
\ket{\downarrow \uparrow}\Big) -
 \xi \Big(\ket{d_L} + \ket{d_R}\Big), \\ 
\ket{s_1} &=& \xi \Big(\ket{\uparrow \downarrow} -
\ket{\downarrow \uparrow}\Big)+ \xi' \Big(\ket{d_L} +
\ket{d_R}\Big),\\ 
\ket{s_2} & = & \frac{1}{\sqrt{2}} \Big(\ket{d_L} -
\ket{d_R}\Big),
\end{eqnarray}
\end{subequations}
where $\ket{\sigma \sigma'}= a^\dagger_{L\sigma}a^\dagger_{R\sigma'}
\ket{0}$ denote singly occupied states, and 
$\ket{d_{i}} = a^\dagger_{i\uparrow}a^\dagger_{i\downarrow}
\ket{0}$, $i=L,R$, are doubly occupied states. The parameters are equal to $\xi =
1/\sqrt{2} \sin (\theta/2)$ and $\xi' =
1/\sqrt{2} \cos (\theta/2)$, where $\theta
= \mathrm{atan}(4\Delta/U)$. The triplet 
states are degenerate in zero magnetic field, and their
eigenenergy is chosen as zero. The singlet state eigenenergies are then,
respectively, $-J, U, U+J$, where $J = 1/2(\sqrt{U^2 + 16
  \Delta^2} - U)$ is the exchange splitting between lowest 
energy singlet and triplet states (see Fig.~\ref{fig:levels}).

The second term describes the QPC leads\cite{gurvitz96,stace04,barrett06,goan01},
\begin{equation}\label{equ:HamilLeads}
H_{{\mathrm{leads}}} = \hbar \sum_{p,\sigma}
\omega_{Sp}a_{Sp\sigma}^\dagger a_{Sp\sigma} + \hbar \sum_{p,\sigma}
\omega_{Dp} a^{\dagger}_{Dp\sigma}a_{Dp\sigma},
\end{equation}
where $\hbar\omega_{np}$ is the energy of an electron in lead $n =
\mathrm{S,D}$ (source, drain) 
and in mode $p$,
$a_{np\sigma}$, $a_{np\sigma}^{\dagger}$ are the corresponding electron
annihilation and creation operators with the additional distinction
of spin $\sigma$.
The third term describes the tunneling of electrons through the QPC\cite{stace04,barrett06,goan01},
\begin{equation}\label{equ:HamTun}
H_{{\mathrm{tun}}} = 
\sum_{p,q,\sigma} (T_{pq} + \chi_{pq}
n_R ) a_{Sp\sigma}^\dagger a_{Dq\sigma} + \mbox{H.c.}
\end{equation}
It consists of two parts: electron tunneling independent of the DQD is
described by the constants $T_{pq}$, while $\chi_{pq}$ quantifies
the tunneling dependent on the Coulomb interaction of QPC electrons
with electrons in the DQD. This depends on the total number of
electrons in the right dot, $n_{\mathrm{R}} = n_{\mathrm{R}\uparrow} +n_{\mathrm{R}\downarrow}$.
The tunneling constants are assumed to be slowly varying over
the energy range where tunneling is allowed\cite{gurvitz96,barrett06},
hence we make the assumption $T_{pq}\approx T$
and $\chi_{pq}\approx \chi$. We assume that the QPC operates in the
high bias regime, that is, the chemical potential offset between the
leads is large enough to induce transitions to doubly excited states
\cite{barrett06}. 

The last two terms in the Hamiltonian describe the energy of the free phonons,
\begin{equation}\label{equ:HamiltPh}
H_{{\mathrm{ph}}} = \sum_{\bm{k},\lambda} 
\hbar\omega_{\bm{k},\lambda} b_{\bm{k},\lambda}^\dagger b_{\bm{k},\lambda},
\end{equation}
and the interaction between phonons and electrons confined in the 
DQD\cite{krummheuer02,vagov02a,vagov03,grodecka05a,roszak10},
\begin{equation}\label{equ:HamiltEPH}
H_{{\mathrm{e-ph}}} = \sum_{\sigma,i} \sum_{\bm{k} ,\lambda}
F_i^{(\lambda)} (\bm{k}) a_{i\sigma}^{\dagger} a_{i\sigma}
(b_{\bm{k},\lambda} + b_{-\bm{k},\lambda}^\dagger).
\end{equation}
In Eqs (\ref{equ:HamiltPh}) and (\ref{equ:HamiltEPH}), 
$b_{\bm{k},\lambda}$ and $b_{\bm{k},\lambda}^{\dagger}$
are phonon annihilation and creation operators for a phonon from
branch $\lambda$ with wave vector $\bm{k}$, 
$\hbar\omega_{\bm{k},\lambda}$ are the corresponding energies,
$
F_{\mathrm{L/R}}^{(\lambda)} (\bm{k}) = F^{(\lambda)}(\bm{k})
e^{\pm ik_x D/2}
$
are electron-phonon coupling constants, and $D$ is the inter-dot distance. We include
deformation potential and piezoelectric couplings.
The coupling constants for the longitudinal ($\lambda = l$) and
transverse ($\lambda = t_{1,2}$) acoustic phonon branches are 
\cite{mahan00,mahan72,roszak09,roszak05b},
\begin{equation}\label{equ:Fl}
F^{(l)}(\bm{k})= \sqrt{\frac{\hbar}{2 \rho_c v
    \omega_{\bm{k},l}}}\left[\sigma k - i \frac{d e}{\varepsilon_0
    \varepsilon_s} M_l (\hat{\bm{k}})\right]\mathcal{F}(\bm{k})
\end{equation}
and
\begin{equation}\label{equ:Ft}
F^{(t_1,t_2)}(\bm{k})= -i \sqrt{\frac{\hbar}{2 \rho_c v
    \omega_{k,t}}} \frac{d e}{\varepsilon_0
    \varepsilon_s} M_{t_1,t_2} (\hat{\bm{k}})\mathcal{F}(\bm{k}),
\end{equation}
respectively,
where $e$ denotes the electron charge, $\rho_c$ is the crystal density,
$v$ is the normalization volume for the phonon modes, $d$ is the
piezoelectric constant, $\varepsilon_0$ is the vacuum permittivity,
$\varepsilon_s$ is the 
static relative dielectric constant, and $\sigma$ is the deformation
potential constant. The functions $M_\lambda$ depend on the
orientation of the phonon wave vector. For the zinc-blende structure
they are given by\cite{mahan72}
\[
M_\lambda (\hat{\bm{k}}) = 2 [
  \hat{k}_x\hat{k}_y(\hat{e}_{\lambda,\bm{k}})_z
  +\hat{k}_y\hat{k}_z(\hat{e}_{\lambda,\bm{k}})_x
+\hat{k}_z\hat{k}_x(\hat{e}_{\lambda,\bm{k}})_y],
\]
where $\hat{\bm{k}} = \bm{k}/k$ and
$\hat{e}_{\lambda,\bm{k}}$ are unit polarization vectors. The form factors
$\mathcal{F}(\bm{k})$ depend on wave-function geometry and are
given by
\[
\mathcal{F}(\bm{k}) = \int \mathrm{d}^3 \bm{r} \psi^*
(\bm{r})e^{i\bm{k}\cdot \bm{r}} \psi
(\bm{r}),
\]
where $\psi(\bm{r})$ is the envelope wave function of an electron
centered at $\bm{r} = 0$.

\section{\label{LindMeq}Lindblad Master equation}
To describe DQD dynamics averaged over a large number of repetitions of the measurement procedure
it is convenient to use the quantum master equation (QME) approach in
Lindblad form (Markov approximation). The problem is relatively
involved due to the interaction of our system of interest (the DQD)
with two reservoirs, a bosonic and  a fermionic one. Since we assume
that these reservoirs are uncorrelated with each other, they can be
treated separately, yielding unconvoluted terms in the QME\cite{breuer02}, 
\begin{eqnarray}\label{equ:lindbladequation}
  \dot \rho (t)& = &- \frac{i}{\hbar} [H_{{\mathrm{DQD}}},\rho] 
\\ \nonumber & &+ 
\frac{1}{\hbar^2} \left (\sum_i^3 C_i\rho C_i^\dagger - \sum_i^3 \frac{1}{2}(C_i^\dagger C_i \rho + \rho
C_i^\dagger C_i )   \right)
 \\  & &+ \frac{1}{\hbar^2} \left(\sum_i^4 B_i\rho B_i^\dagger - \sum_i^4 \frac{1}{2}(B_i^\dagger B_i \rho + \rho
 B_i^\dagger B_i ) \right).\nonumber
\end{eqnarray}
Here, the first term on the right side of the equation describes the
free DQD evolution, while the Lindblad operators  $C_i$ relate to the
DQD-QPC interaction, and $B_i$ describe phonon-related
effects. 

The Lindblad operators $C_i$ may be
obtained following Ref. [\onlinecite{barrett06}]
in the Born-Markov and rotating wave approximations (RWA)
and assuming independence of the tunneling rates on the initial and
final electron state within the relevant energy regime.
The operators are of the form\cite{barrett06}
\begin{eqnarray*}
  C_1 & = & \nu \sqrt{\frac{V-(U+J)}{\hbar}} \sin\frac{\theta}{2} \ketbra{s_2}{s_0}, \\
  C_2 & = & \nu \sqrt{\frac{V+(U+J)}{\hbar}} \sin\frac{\theta}{2} \ketbra{s_0}{s_2}, \\
  C_3 & = & \sqrt{\frac{V}{\hbar}} \left[ (\mathcal{T} + \nu)\mathbb{I} +\nu \cos\frac{\theta}{2} 
\left(\ketbra{s_1}{s_2} + \ketbra{s_2}{s_1}\right) \right],
\end{eqnarray*}
where $V = (\mu_S - \mu_D)$ is the QPC bias, 
$\mathcal{T}=\sqrt{4\pi g_Lg_R}T$ is the unconditional tunneling constant
related to $T_{pq}$ of Eq. (\ref{equ:HamTun}) and $\nu=\sqrt{4\pi
  g_Lg_R}$ is a constant 
stemming from tunneling conditioned on the occupation of the right QD
(related to $\chi_{pq}$), where $g_i$ is the density of states of the $i$-th lead ($i=L,R$).
$C_1$ and $C_2$ describe inelastic
transitions which involve energy transfer between the DQD and QPC
electrons accompanied by transitions between the low energy state
$\ket{s_0}$ and high energy states (blue arrows in Fig \ref{fig:levels}). The quasi-elastic transition
between states of similar energy $\ket{s_1}$ and $\ket{s_2}$ is
represented by the Lindblad operator $C_3$; this operator also describes the fully
elastic processes corresponding to electrons tunneling through the QPC
without disturbing the DQD state (which are also possible in a spin-triplet DQD state).

To describe the electron-phonon interaction it is convenient to
rewrite the appropriate Hamiltonian (Eq. (\ref{equ:HamiltEPH})) in the
basis of DQD eigenstates\cite{roszak09}, Eqs.~(\ref{eqn:states}),
\begin{eqnarray*}
 H_{\mathrm{e-ph}} & = &\sqrt{2}\left[\xi'
(\ketbra{s_1}{s_2} + \ketbra{s_2}{s_1})\right.\\
&&\left.- \xi(\ketbra{s_0}{s_2} +
 \ketbra{s_2}{s_0}) 
\right](\widehat{F}_L -
 \widehat{F}_R),
\end{eqnarray*}
where the operators are $\widehat{F}_{\mathrm{L/R}} =
\sum_{\bm{k},\lambda} F^{(\lambda)}_{\mathrm{L/R}} (\bm{k}) 
(b_{\bm{k},\lambda} + b_{-\bm{k},\lambda}^\dagger)$.

Following the standard method\cite{breuer02} we obtain the phonon Lindblad
operators in the Born-Markov and RWA approximations (schematically represented by red arrows in Fig.~\ref{fig:levels}),
\begin{eqnarray*}
  B_1 & = & \sqrt{\gamma_{02}} \ketbra{s_0}{s_2}, \\ 
  B_2 & = & \sqrt{\gamma_{20}} \ketbra{s_2}{s_0} = \sqrt{\gamma_{02}  e^{-(U+J)/k_B T}} \ketbra{s_2}{s_0}, \\
  B_3 & = & \sqrt{\gamma_{12}} \ketbra{s_1}{s_2} = \sqrt{\gamma_{21} ~ e^{-J/k_B T}} \ketbra{s_1}{s_2}, \\
  B_4 & = & \sqrt{\gamma_{21}} \ketbra{s_2}{s_1},
\end{eqnarray*}
where the transition rates are $\gamma_{ij} = 2\pi R_{ij}(\omega_{ij})$,
with the relevant spectral densities of the phonon reservoir defined
as\cite{roszak05b} 
\begin{eqnarray}\label{equ:R02}
R_{02}(\omega) &= &\frac{8 \xi^2}{N} 
\sum_{\bm{k},\lambda} |F^{(\lambda)} (\bm{k})|^2 \sin^2
\frac{k_xD}{2} \\
& & \times \left[ (n_{\bm{k}} +1) \delta(\omega- \omega_k) + n_{\bm{k}} \delta(\omega + \omega_k)\right],\nonumber
\end{eqnarray}
\begin{eqnarray}\label{equ:R21}
R_{21}(\omega) &= &\frac{8 \xi'^{2}}{N} 
\sum_{\bm{k},\lambda} |F^{(\lambda)} (\bm{k})|^2 \sin^2
\frac{k_x D}{2}  \\
& & \times \left[ (n_{\bm{k}}+1) \delta(\omega- \omega_k) + n_{\bm{k}} \delta(\omega + \omega_k)\right],\nonumber
\end{eqnarray}
where $n_{\bm{k}}$ denotes the
Bose distribution and $R_{ij}(\omega)=R_{ji}(\omega)$.
The energy differences between the states are equal to $\hbar
\omega_{02} =- \hbar \omega_{20} = U+J$ and $\hbar \omega_{21} = -
\hbar \omega_{12} = J$.
Note that phonon related processes involve exactly the same
pairs of singlet states as transitions related to the QPC (see Fig.~\ref{fig:levels}).
In the zero-temperature limit, $\gamma_{20} = \gamma_{12} = 0$.

Solving the QME given by Eq.~(\ref{equ:lindbladequation})
in the long time limit results in distinct steady states in the spin-singlet subspace
and the spin-triplet subspace.
The triplet steady state $\rho_{\infty}^{(\mathrm{t})}$ can be any
superposition of the triplet states. 
The singlet steady state is equal to
\begin{eqnarray}
\label{ssss}
\rho_\infty^{(\mathrm{s})} &=& \frac{1}{N}
\left[(A_+^2 + \gamma_{02})(A_0^2 + \gamma_{21})\ketbra{s_0}{s_0}
  \right .\\ & &
+ (A_0^2 + \gamma_{12})(A_-^2 + \gamma_{20})\ketbra{s_1}{s_1}
\nonumber  \\ & &+
\left . (A_-^2 + \gamma_{20})(A_0^2 + \gamma_{21})\ketbra{s_2}{s_2}\right ],  \nonumber 
\end{eqnarray}
where 
\begin{eqnarray}
\nonumber
N &=& (A_+^2 + \gamma_{02})(A_0^2 +\gamma_{21}) + (A_0^2
+\gamma_{12})(A_-^2 + \gamma_{20}) \\
\nonumber
&&+(A_-^2 + \gamma_{20})(A_0^2 + \gamma_{21}),\\
\nonumber
A_{\pm} &=& \nu \sqrt{(V\pm J \pm U)/\hbar} \sin (\theta/2),\\
\nonumber
A_0 &=&\nu  \sqrt{V/\hbar} \cos (\theta/2).
\end{eqnarray}
Hence, the singlet steady state depends explicitly on the strengths of the phonon-couplings
relative to the QPC coupling strengths.

\section{\label{sec:Method}The method of stochastic simulation}

Modern experimental techniques allow one to observe single
system evolutions\cite{bylander05,cassidy07,gustavsson08} which cannot be reproduced by the QME
approach. In our system, this entails the situation when, beginning
with the same spin singlet-triplet superposition state, the DQD system
may end up in either the triplet state or a singlet-only
superposition due to the measurement executed by the
QPC. Furthermore, the evolution between initial and final states is
prone to variation on different runs even if the two states are
always the same. To
describe such a physical process probabilistic elements need to be
introduced into the
evolution
\cite{meystre07,breuer02,goan01,goan01a,stace04,barrett06,wiseman01}.  

To model a single measurement run one introduces a
conditional density matrix (state) $\rho_{\mathrm{c}}(t)$ that depends
on the history 
of counting events (corresponding to electron tunnelings through the
QPC in our case) and a counting process $N(t)$, corresponding
to the number of electrons that have passed through the QPC. 
The stochastic equation describing the conditional state (in the
interaction picture) has the form\cite{goan01a}
\begin{eqnarray}
\rho_{\mathrm{c}}(t+dt) & = & 
\frac{\rho_{\mathrm{c}}-(i/\hbar)dt[H_{\mathrm{eff}},\rho_{\mathrm{c}}]_{*}
+\mathcal{L}_{\mathrm{ph}}[\rho_{\mathrm{c}}]dt}{
1-P_{1}}\left( 1-dN \right)
\nonumber \\
&&
+\frac{\sum_{i }C_{i}\rho_{\mathrm{c}} C_{i}^{\dag}}{P_{1}} dN,
\label{SDE}
\end{eqnarray}
where $[H_{\mathrm{eff}},\rho]_{*}
=H_{\mathrm{eff}}\rho-\rho H_{\mathrm{eff}}^{\dag}$, 
with the non-hermitian operator 
\begin{displaymath}
H_{\mathrm{eff}}=-\frac{i\hbar}{2} \sum_i C_i^\dagger C_i,
\end{displaymath}
and
\begin{displaymath}
\mathcal{L}_{\mathrm{ph}}[\rho] = \frac{1}{\hbar^2} \left[ \sum_i
B_i\rho B_i^\dagger - \sum_i
\frac{1}{2}(B_i^\dagger B_i \rho + \rho B_i^\dagger B_i ) \right]
\end{displaymath}
is the Lindblad generator accounting for the electron-phonon
interaction. The increment of the counting process, $dN$, can be zero or one,
depending on whether an electron tunneling event was observed in the
time interval $dt$. The statistics of this increment is defined by the
probability of a tunneling event (conditional on the history of the
measurement events)
$P_{1}=P_{\mathrm{c}}[dN=1]=\sum_{i} \tr C_{i}^{\dag}C_{i}\rho_{c} dt$.
The dissipative contribution described by $\mathcal{L}_{\mathrm{ph}}$  
appears as a result of
averaging over the unobserved phonon scattering events and corresponds
to the fact that the conditional density matrix describes only one
subsystem of the interacting carrier-phonon system. 
Some of the terms proportional to $dNdt$ (that could formally be
omitted to 
obtain the equation in its most common form\cite{
  goan01,goan01a,stace04,barrett06,wiseman01}) are 
kept in Eq.~\eqref{SDE} for clarity of interpretation.

Thus, the first term in Eq.~\eqref{SDE} describes the evolution in the
absence of electron tunneling events. This evolution is described by
the deterministic equation
\begin{eqnarray*}
\left.\dot{\rho_{\mathrm{c}}}(t)\right|_{\mathrm{det}} & = &
\frac{\rho_{\mathrm{c}}-(i/\hbar)[H_{\mathrm{eff}},\rho_{\mathrm{c}}]_{*}
+\mathcal{L}_{\mathrm{ph}}[\rho_{\mathrm{c}}]}{1-P_{1}}-\rho\\
&\approx& -\frac{i}{\hbar}[H_{\mathrm{eff}},\rho_{\mathrm{c}}]_{*}
+\mathcal{L}_{\mathrm{ph}}[\rho_{\mathrm{c}}]+ 
\sum_{i} \tr C_{i}^{\dag}C_{i}\rho_{\mathrm{c}}
\end{eqnarray*}
(neglecting terms on the order of $(dt)^{2}$).
This equation is satisfied by 
$\rho_{\mathrm{c}}=\tilde{\rho_{\mathrm{c}}}/(\tr \tilde{\rho}_{\mathrm{c}})$,
where the unnormalized conditional density matrix
$\tilde{\rho}_{\mathrm{c}}$ evolves according to
\begin{equation}\label{determ}
\left.\dot{\tilde{\rho_{c}}}(t)\right|_{\mathrm{det}} 
= -\frac{i}{\hbar}[H_{\mathrm{eff}},\tilde{\rho}_{\mathrm{c}}]_{*}
+\mathcal{L}_{\mathrm{ph}}[\tilde{\rho}_{\mathrm{c}}].
\end{equation}
Clearly, the trace
of $\rho_{\mathrm{c}}$ decreases under the evolution described by
Eq.~\eqref{determ},
\begin{equation}\label{trace}
\frac{d}{dt}\tr\tilde{\rho}_{\mathrm{c}}(t)=
-P_{1}(t) \tr\tilde{\rho}_{\mathrm{c}}(t).
\end{equation}
The second term in Eq.~\eqref{SDE},
which contributes if a tunneling event has taken place ($dN=1$),
corresponds to a discontinuous change of the system state (a
\textit{jump}). 

The conditional density matrix thus follows the continuous evolution
described by Eq.~\eqref{determ} interrupted by jumps corresponding to
electron flow through the QPC. 
In order to simulate this piecewise continuous evolution we
generalize the method of finding the cumulative distribution function
$F(t)$ for the random jump time $t$ proposed for stochastic wave function
simulations\cite{breuer02,gardiner04}. We note that the survival
probability $\tilde{F}(t)=1-F(t)$ satisfies the same Eq.~\eqref{trace}
as $\tr \tilde{\rho}_{\mathrm{c}}$. Both these quantities are equal to 1 at the
initial time of the deterministic evolution interval. Hence,
$F(t)=1-\tr \tilde{\rho}_{\mathrm{c}}$. Based on this, the conditional evolution can be
simulated by solving the deterministic equation for the unnormalized
conditional density matrix (which, in this case, can in principle be
done analytically) and finding the next jump time by generating a
random number according to the known cumulative distribution function.

\section{\label{sec:Results}Results}

The stochastic method presented in the previous section allows us to
model single realizations of the evolution
and subsequently to analyze the
current flowing through the QPC in a given realization. 
In the absence of phonons,
the noise characteristics of this current may serve to distinguish the
spin-singlet and spin-triplet DQD states\cite{barrett06}, hence, the
interaction between QPC and dot electrons can be regarded as a
measurement of two-electron DQD spin states by the QPC. 
Since we are dealing with a solid state system, one can expect 
that phonon-related effects will disturb this measurement
process.

\begin{figure}[tb]
  \includegraphics{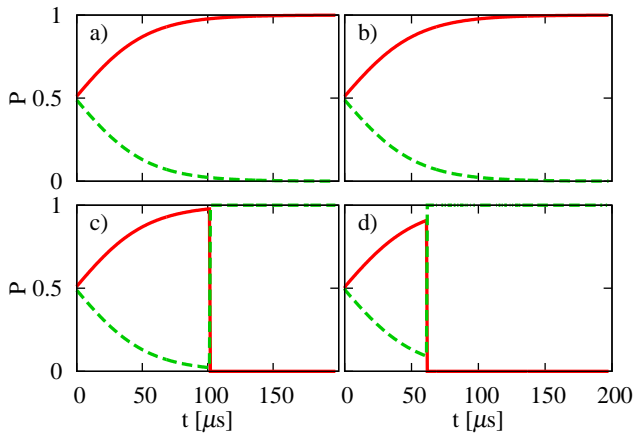}
  \caption{Exemplary singlet (green, dashed) and triplet (red, solid)
    probability time-evolutions for final triplet (a, b) and
    singlet (c, d) states, without (a, c) and with the
    phonon interaction (b, d).}  
  \label{Fig:Fig_both_singlet_tryplet}
\end{figure}

%\begin{widetext}
\begin{figure*}[tb]
  \includegraphics[scale=1.085]{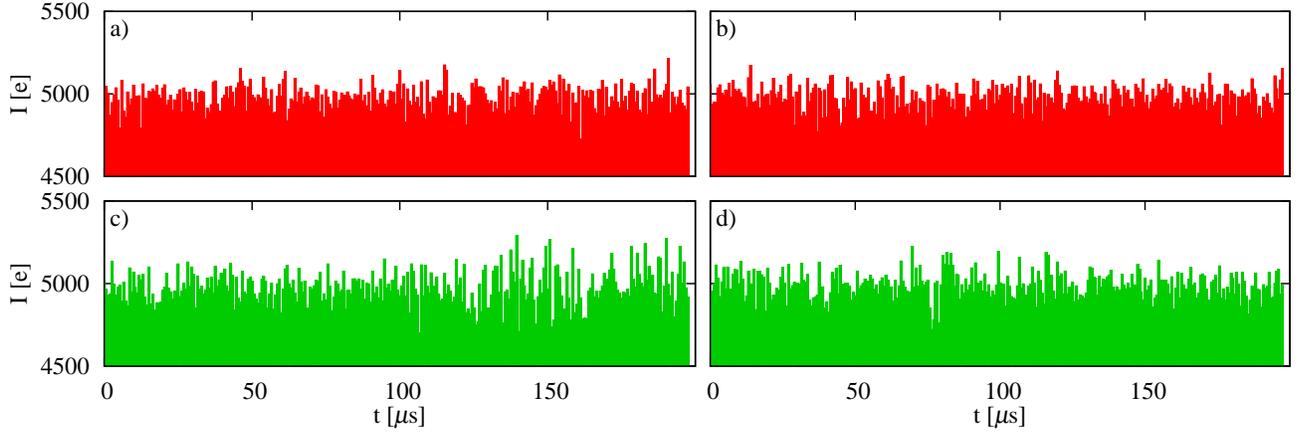}
  \caption{QPC current as a histogram of tunneling events for triplet (a, b) and
    singlet (c, d) states, without (a, c) and with the
    phonon interaction (b, d). The time interval used for the
    histogram is 0.66~$\mu$s.}  
  \label{fig:histogramPHNOPH}
\end{figure*}
%\end{widetext}

In the following, parameters corresponding to GaAs structures
are used\cite{barrett06,loss98}. 
The DQD energies are $U = 1$ meV and $J = 0.1$ meV, and the QPC bias is taken to be
$V = 2$ meV. 
Unless stated otherwise, the QPC tunneling parameters are fixed at 
$\mathcal{T} =4\cdot 10^{-2}$ and $\nu = 9 \cdot
10^{-4}$.
The material
parameters relevant for the calculation of the spectral density of the
phonon reservoir are\cite{roszak09} $c_L = 5100$ m/s, $c_t = 2800$
 m/s, $\epsilon_s = 13.2$, $d = 0.16$ C/m$^2$, $\sigma
= -8.0$ eV and $\rho_c = 5369$ kg/m$^3$. Two-dimensional Gaussian single-electron wave
functions were used with
170 nm full width at half maximum of the probability density, and the
distance between the dots was set to $D = 250$ nm. This yielded
zero-temperature phonon transition rates $\gamma_{02} = 1.15 \cdot
10^{-3}$ ns$^{-1}$ and
$\gamma_{21} = 6.01 \cdot 10^{-8}$ ns$^{-1}$ ($\gamma_{20} = \gamma_{12}=0$).
The choice of QPC parameters corresponds to the phonon interaction being roughly
$2.5$ times stronger than the QPC interaction, meaning that 
$\sqrt{\gamma_{02}}=2.5 \nu\sqrt{V/\hbar}$.

The results presented in this and the following Section 
are all taken in the zero temperature limit. This is because
experimental realizations of QPC measurements are performed at temperatures 
that do not exceed $0.1$ K, leading to 
extremely low phonon transition rates from lower to higher energy states.

Fig.~\ref{Fig:Fig_both_singlet_tryplet} shows exemplary time evolutions of the
probability of finding the DQD in a spin-singlet state (green, dashed
lines) or a spin-triplet state (red, solid) for an initial equal superposition state
\begin{equation}
\label{equal}
\ket{\Psi} = \frac{1}{\sqrt{2}}(\ket{s} + \ket{t}), 
\end{equation}
where $\ket{t}$
can be any superposition of the triplet states, Eqs (\ref{trypleta})-(\ref{trypletc}), and 
\begin{equation}
\ket{s}  =
\cos\frac{\theta}{2} \ket{s_0} + \sin \frac{\theta}{2}\ket{s_1}=
\frac{1}{\sqrt{2}}(\ket{\uparrow\downarrow} - \ket{\downarrow\uparrow}),
\end{equation}
as in Ref. [\onlinecite{barrett06}].
The top panels show instances where the final state is a spin-triplet
(the measurement outcome was the triplet state),
while the bottom-panel evolutions ended up in the spin-singlet state
(the measurement outcome was the singlet state).
The electron-phonon interaction is
included only in the right panels; the phonon influence on coherence and
localization is studied in more detail later on.
As seen, regardless of the presence of the electron-phonon coupling, the continuous evolution leads 
to DQD localization in the triplet 
spin state (see Fig.~\ref{Fig:Fig_both_singlet_tryplet} a, b),
while for there to be a measurement of a singlet spin state,
the occurrence of a quantum jump is required. 

The simulation results for the QPC currents corresponding to the final
singlet and triplet states are shown in 
Fig.~\ref{fig:histogramPHNOPH}. Even through the evolutions depicted in 
Fig.~\ref{Fig:Fig_both_singlet_tryplet} show no clear difference between the phonon and no-phonon
cases, in the phonon-free situation in Fig.~\ref{fig:histogramPHNOPH} (left panels)
a difference in the magnitude of the current fluctuations (noise) can be seen
between the singlet and triplet case, while no such distinction is evident in the current
when the phonon influence is included. 
For the realistic choice of material parameters, QPC and DQD properties, and for our choice of 
counting time step,
the differences are relatively small, but still a period of time (after about $110$ microseconds)
when the DQD electrons occupy higher energy singlet states resulting in 
increased current fluctuations
can be seen (Fig.~\ref{fig:histogramPHNOPH} c).
When the phonon coupling is included (Fig.~\ref{fig:histogramPHNOPH} b, d)
this distinction is diminished (to the level that no time period of increased fluctuations
can be seen with the ``bare eye''), so the measurement effect
is suppressed. 

Clearly, such observations based on the informal analysis of the current noise trace
are to a large extent subjective and cannot form the base for rigorous conclusions
on the measurement outcome or for assessing the role of phonon-induced dissipation.
In order to provide a firm ground for such a discussion,
the quantitative noise characteristics, needed to fully
describe phonon influence 
on the distinguishability of the spin-singlet and spin-triplet states in the QPC spin
measurement setup, are studied in the next section.

\begin{figure}[tb]
  \includegraphics[width=85mm]{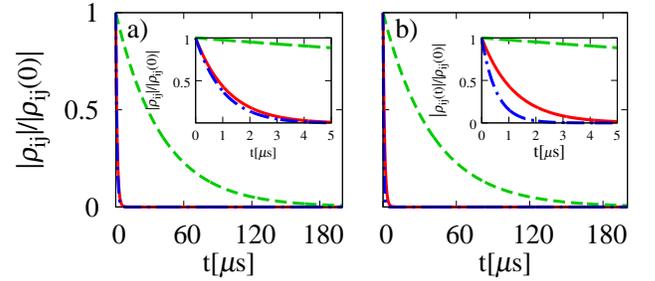}
  \caption{Normalized singlet-triplet coherences without
    phonon influence (a) and with phonon influence (b); 
    $|\dmatrix{t}{\rho}{s_0}|$: red solid lines,
    $|\dmatrix{t}{\rho}{s_1}|$: green dashed lines, and $|\dmatrix{t}{\rho}{s_2}|$: 
	blue dash-dotted lines. Inset: short-time evolution.}  
  \label{fig:dek}
\end{figure}

The most obvious effect of coupling to phonons can be seen in the 
dynamics of coherence
between singlet and triplet states expressed by the normalized amplitudes of the 
off-diagonal 
density matrix elements averaged over many realizations of single measurement
simulations. The coherence time between the triplet and the
$|s_2\rangle$ singlet states, 
$\tau_{s_2}=40.62$ $\mu$s, 
is over an order of magnitude longer than the other two coherence times,
which are $\tau_{s_0}=1.03$ $\mu$s for $|s_0\rangle$ and $T_{s_1}=1.20$ $\mu$s
for $|s_1\rangle$.
Furthermore, phonon influence (Fig.~\ref{fig:dek} b)
also strongly varies depending on the particular coherence in question,
leaving the triplet $|s_1\rangle$ singlet coherence time unchanged,
slightly influencing the long
$|s_2\rangle$ coherence, the time of which is now $\tau_{s_2}=40.55$ $\mu$s,
and cutting the coherence time of $|s_0\rangle$ almost by half, leaving
$\tau_{s_2}=0.54$ $\mu$s. This is because different phonon transition
rates $\gamma_{ij}$ 
govern the evolution of different coherences, and the rates significantly
differ from one another. The coherence between a triplet and the
$|s_0\rangle$ singlet 
is influenced both by the strong transition from $|s_2\rangle$ to $|s_0\rangle$
and the weaker transition from $|s_1\rangle$ to $|s_2\rangle$, hence 
phonon influence on the coherence times is big.
The triplet--$|s_1\rangle$ coherence
is only related to transitions from $|s_0\rangle$ to $|s_2\rangle$
and $|s_2\rangle$ to $|s_1\rangle$ which are very weak at low temperatures
and do not occur at zero temperature, while
the $|s_2\rangle$ coherence depends on the transition rates 
from $|s_1\rangle$ to $|s_2\rangle$
and $|s_0\rangle$ to $|s_2\rangle$, of which the first is small and the second is negligible
in the low temperature range.

\begin{figure}[tb]
  \includegraphics[width=65mm]{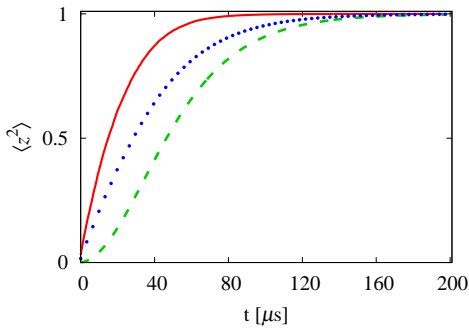}
  \caption{Localization of the DQD state with singlet post-selection (red solid line), 
triplet post-selection (green dashed) and with no state post-selection (blue dotted).}  
  \label{fig:loc}
\end{figure}

While the dephasing rates are the most common characteristics of open
system dynamics, from the point of view of the measurement process the
time of localization into  one of  the measurement basis states (the
``collapse'' of the system state) is of more interest. To quantify how fast 
the DQD system  
reaches the singlet or triplet subspaces, we introduce a measure of localization
analogous to the degree of localization used in the description of
charge measurement \cite{goan01a}. It determines the timescales on
which the QPC measurement 
on the DQD spin states
is performed. For the singlet-triplet measurement, the observable
quantifying the localization 
is $\langle z^{2} \rangle $, where 
\begin{displaymath}
z= \sum_{i} \langle t_{i}|\rho_{\mathrm{c}}|t_{i} \rangle -
\sum_{i} \langle s_{i}|\rho_{\mathrm{c}}|s_{i}\rangle 
\end{displaymath}
and the averaging is performed over many simulated measurement runs.
The quantity is equal to zero when the occupation of the singlet and triplet subspaces is equal
and grows with the absolute value of the difference between the two occupations,
reaching unity when the DQD state is either a fully triplet or a fully singlet state.
Since the singlet subspace is the orthogonal complement of the triplet
subspace, the degree of localization 
can be described using the probability of finding the system in the triplet state, which gives
$z=1-2\langle t|\rho|t \rangle $.
The blue dotted line in Fig.~\ref{fig:loc} shows the localization, $\langle z^2(t)\rangle$,
for the initial equal superposition state [Eq. (\ref{equal})]
for which $\langle z^2(0)\rangle=0$.
The red solid line and green dashed lines depict the evolution of localization
for the same set of data, on which a post-selection into the set of measurement
runs that yielded a singlet outcome (red solid) and the set that yielded a triplet outcome
(green dashed) was performed. Although the localization occurs on the scale of tens of microseconds
regardless of the measurement outcome, the localization 
is significantly faster when a singlet is measured, 
with localization time $\tau_{\mathrm{loc}}^{(\mathrm{s})} = 20$
$\mu$s, than when triplet is measured, 
with localization time $\tau_{\mathrm{loc}}^{(\mathrm{t})} =46$ $\mu$s. 
This is because a quantum jump is needed to localize in the spin-singlet, and
the occurrence of jumps shortens the localization time (as seen in 
Fig.~\ref{Fig:Fig_both_singlet_tryplet}). 
If no post-selection is made then 
localization time is $\overline{\tau}_{\mathrm{loc}} = 35$ $\mu$s.
As expected\cite{marcinowski11}, phonons have no effect on the localization times
as long as they are not monitored (meaning that no measurement is performed
on the phonon subsystem). 

As can be seen from the above discussion, the phonon-induced
dissipation destroys coherence but does not change the localization
time, which depends only on the coupling to the measurement device. 
The major phonon effect, from the point of view of the measurement, is
the reduction of QPC current noise in the DQD singlet state  
due to the phonon induced suppression of transitions to the
excited singlet states which are responsible for
the increased singlet current noise. 

\section{\label{sec:Noise}Noise characteristics}

%\begin{widetext}
\begin{figure*}[tb]
  \includegraphics[scale=0.95]{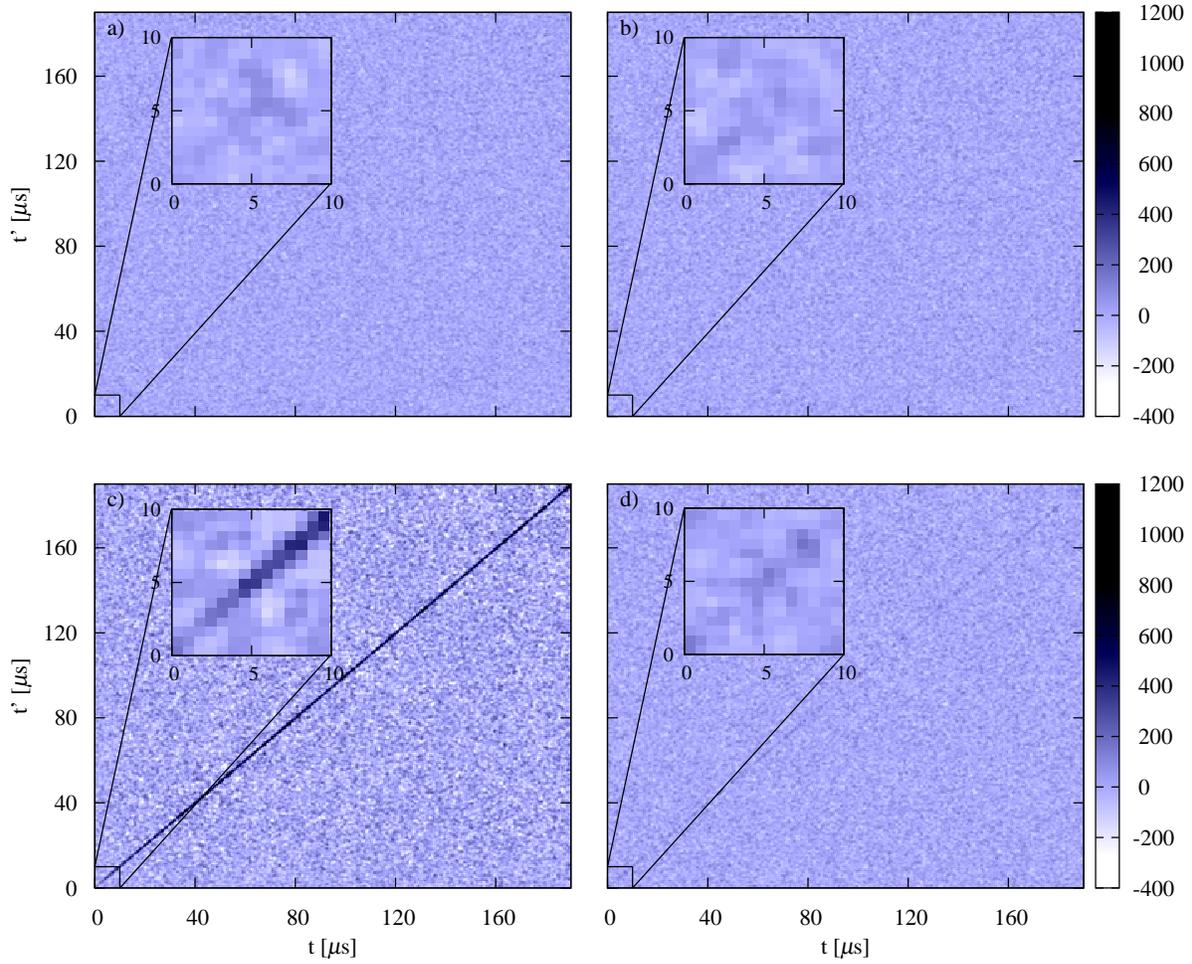}
  \caption{Current-current correlation function $G(t,t')-G(t=t')$ 
of initial equal superposition state
for the triplet (a, b)
and singlet (c, d) measurement outcomes without (a, c) and with the phonon
interaction (b, d). Insets: Detail at short times.}  
  \label{fig:mapa}
\end{figure*}
%\end{widetext}

Although the destructive phonon effects might already be seen in the QPC
current noise of single evolutions, a general noise characteristic is much
more revealing\cite{blanter01}. 
Fig.~\ref{fig:mapa} shows the two-time maps of the current-current correlation
function\cite{blanter01,barrett06,goan01,stace04},
\begin{equation}
G(t,t') = \langle I(t') I(t)\rangle - \langle I(t')\rangle
 \langle I(t)\rangle,
\label{cc}
\end{equation}
where the average $\langle ...\rangle$ is taken over $\sim 5000$
measurement runs (the small-scale fluctuations in the maps are due to
the finite number 
of runs). The strongly peaked values of the current-current
correlation function for $t=t'$ 
are not shown on the maps for the sake of clarity.
The initial state is chosen to be the
equal superposition state of Eq. (\ref{equal}).
In the top panels (Fig.~\ref{fig:mapa} a, b), 
the correlation function for post-selected spin-triplet measurement outcomes
is shown, while in the bottom panels
(Fig.~\ref{fig:mapa} c, d) the post-selected spin-singlet outcomes are depicted.
Again, the left panels (Fig.~\ref{fig:mapa} a, c) correspond to the no-phonon case and the right panels 
(Fig.~\ref{fig:mapa} b, d)
correspond to zero-temperature phonons.
When the system reaches its steady state, the correlation function no longer depends
on the two times $t$ and $t'$, but is a function only of $\tau=t'-t$.
As can be seen, regardless of the presence of phonons, the QPC current
noise is Poissonian
(meaning that there are no correlations present for $t\neq t'$)
when the DQD is in the triplet state.
In the singlet case, current-current correlations are present for small time differences $\tau$,
leading to the appearance of the diagonal line (of a finite width) on the lower left map.
In the presence of phonons, these correlations are strongly
suppressed due to phonon processes that preclude long intervals of
occupation of excited singlet states, and the 
visibility of the diagonal line is much worse (lower right).

The insets on each map of Fig.~\ref{fig:mapa}
contain an enlargement of the area corresponding to small $t$ and $t'$, 
when the system has not yet localized in either a singlet or
a triplet steady state. On the scale of several microseconds
(consistent with the localization time shown in Fig.~\ref{fig:loc}),
we observe small correlations in the upper panels which then disappear as the 
DQD state approaches the purely triplet state. On the lower panels we
can observe the finite time over which the correlations build up,
corresponding to
a decreased visibility of the correlation feature before the
DQD subsystem localizes in the singlet steady state.

\begin{figure}[ht]
  \includegraphics[scale=1.4]{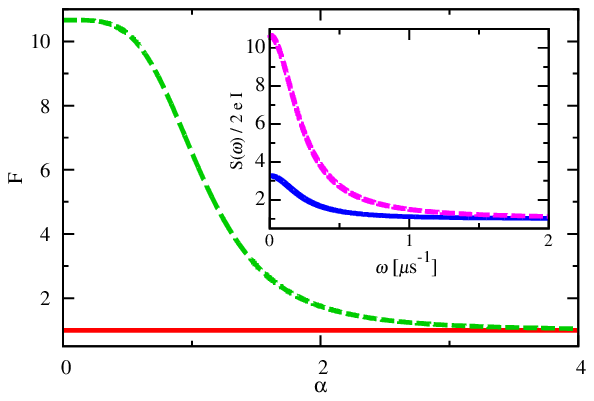}
  \caption{Steady state Fano factor for the QPC current as a function
    of the relative strength of the electron-phonon interaction for
    singlet (green dashed line) and triplet states (red
    solid). Inset: Normalized singlet steady state noise power spectra
    of the QPC detector current without (violet dashed line) and with
    electron-phonon interaction (blue solid), $\alpha=1.5$.
\label{Fig:Fano}}
\end{figure}

To quantify the influence of phonon effects on the measurement scheme
and its dependence on the ratio between the DQD-QPC interaction and 
the electron-phonon coupling we use the Fano factor\cite{fano47,blanter01,flindt10,roszak12}.
It is a widely used noise measure, defined as the zero-frequency shot
noise power normalized to Poissonian shot noise power,
$F={S(0)} / {2e\bar{I}}$, where $\bar{I}$ is the mean current.
The QPC noise power spectrum for steady states is given by $S(\omega) = 2
\int_{\infty}^{\infty} \mathrm{d}\tau G(\tau) e^{-i\omega \tau}$, where
the current-current correlation function, which has been introduced in Eq. (\ref{cc}),
now only depends on $\tau=t'-t$.
Following Refs~[\onlinecite{goan01,barrett06}] we can calculate the
correlation functions 
for singlet and triplet steady states,
\begin{eqnarray}
\nonumber
G^{(t/s)}(\tau) & = &  e^2 \left[\mathrm{Tr} \{\sum_{i,j}
  C_{j} [e^{\mathcal{L}(\tau-\Delta t)} C_i\rho^{(t/s)}_{\infty}C_i^\dagger]
  C_{j}^\dagger\}\right. \\ 
\nonumber
& & - \mathrm{Tr} \{ \sum_i C_i \rho^{(t/s)}_\infty
  C_i^\dagger\}^2\\ 
&& + \left.\mathrm{Tr} \{ \sum_i C_i \rho^{(t/s)}_\infty
  C_i^\dagger\} \delta(\tau)\right]. 
\end{eqnarray}
Spectra for the
different measurement outcomes can be found by substituting the appropriate
steady states: the singlet state is given by Eq. (\ref{ssss}) and
$\rho_{\infty}^{(t)}=|t\rangle\langle t|$. 

Fig.~\ref{Fig:Fano} shows the singlet and triplet
Fano factor curves as a function of the relative coupling strengths
of the DQD to the phonon reservoir and to the QPC.
For the sake of realism it is the tunneling parameters of the QPC
which are changed, while the electron-phonon interaction is kept at a value
corresponding to realistic gate defined QDs (see end of Section \ref{LindMeq}
and beginning of Section \ref{sec:Results}).
The scaling parameter $\alpha= \frac{\mathcal{T}_0}{\mathcal{T}} =
\frac{\nu_0}{\nu}$, with $\mathcal{T}_0 = 0.1$ and $\nu_0 = 2.25 \cdot
10^{-3}$, is chosen in such way that $\alpha = 1$ corresponds to the
situation when the interaction with phonons is roughly the same
strength as the interaction with the QPC; 
this means that $\sqrt{\gamma_{02}}=\nu_0 \sqrt{V/\hbar}$.
As can be seen, the phonons dominate at large
$\alpha$ (small QPC current) leading to a suppression of the noise difference
and breaking of the measurement scheme, while for large currents
their effect is negligible.
Note that the results discussed earlier in this paper correspond to
the QPC interaction strength $\alpha=2.5$.
Even though the phonon-induced suppression of the spin-singlet Fano factor
at this moderate value of the scaling parameter is small,
the effects seen in Figs \ref{Fig:Fig_both_singlet_tryplet} - \ref{fig:mapa} are already 
non-negligible. 
The normalized singlet steady state noise power as a function of frequency 
is shown in the inset of Fig.~\ref{Fig:Fano}. The blue solid line corresponds to $\alpha=1.5$,
while the violet dashed line depicts the no-phonon situation at the same QPC current
values. As can be seen, phonon-interactions lead to noise damping for
the whole noise power  spectrum.
On the other hand, the width of the noise power spectrum $S(\omega)$
does not change when the phonon effects are included. This obviously
means that the width of the corresponding feature in the correlation
function $G(\tau)$ is insensitive to phonons. Since this width sets
the minimum time scale over which the measurement has to be continued
in order to extract the spectral characteristics shown in
Fig.~\ref{Fig:Fano}, this leads to the conclusion that also
this measurement-related time scale is not affected by phonon-induced
decoherence. 

\section{\label{sec:Conclusions}Conclusions}

We have studied phonon-influence on the QPC measurement of
two-electron DQD spin states, 
relying on different characteristics of the resulting QPC current noise. 
We have shown that although phonons destroy the singlet-triplet
coherence, they  do not affect the localization  
of the DQD into the measurement basis and therefore do not influence
this contribution to the measurement time. Also the time necessary to
establish the spectral properties of the noise, which is on the order
of the correlation time, does not change when carrier-phonon coupling
is included. Nonetheless, phonons do disturb the measurement by
impeding the distinguishability of the spin-singlet and spin-triplet
configurations. This is due to the phonon-induced suppression of singlet-singlet
transitions between low and high energy states which are responsible 
for the differences in the noise observed for the triplet and singlet
spin symmetries. This is reflected, in particular, by the reduced amplitude of the 
singlet-related
feature in the noise power spectrum and the suppressed super-Poissonian character
of the Fano factor.

We have found that the perturbing phonon effects with respect to the measurement
distinguishability
at moderate relative strengths of the DQD-QPC and electron-phonon couplings
are already non-negligible. Furthermore, the electron-phonon interaction
can lead to complete indistinguishability of the two spin configurations, if it is strong
enough compared to the DQD-QPC coupling.
This means that the mechanism may render the measurement scheme
completely useless, 
while it does not impose a lengthening of the times of the measurement-induced 
localization into the triplet and singlet subspaces.

\section{\label{sec:Acknowledgments}Acknowledgments}
We are very grateful to Ryszard Buczko and Jan Mostowski for discussions
which inspired this work, and to Tom\'{a}\v{s} Novotn\'{y} and 
Mateusz Kwa{\'s}nicki for helpful discussions relating to noise 
characteristics. This work was supported in parts by the TEAM programme of 
the Foundation for Polish Science, co-financed from the European Regional 
Development Fund and by the Polish NCN Grant No. 2012/05/B/ST3/02875.
The calculations were carried out in the 
Wroc{\l}aw Centre for Networking and Supercomputing (http://www.wcss.wroc.pl), 
grant no. 203. 

\bibliographystyle{prsty}
\bibliography{abbr,quantumlukasz}

\begin{thebibliography}{10}

\bibitem{loss98}
D. Loss and D.~P. DiVincenzo, Phys. Rev. A {\bf 57},  120  (1998).

\bibitem{nowack07}
K.~C. Nowack, F.~H.~L. Koppens, Y.~V. Nazarov, and L.~M.~K. Vandersypen,
  Science {\bf 318},  1430  (2007).

\bibitem{greilich09}
A. Greilich, S.~E. Economou, S. Spatzek, D.~R. Yakovlev, D. Reuter, A.~D.
  Wieck, T.~L. Reinecke, and M. Bayer, Nature Physysics {\bf 5},  262  (2009).

\bibitem{kim11}
N.~Y. Kim, K. Kusudo, C. Wu, N. Masumoto, A. Loffler, S. Hofling, N. Kumada, L.
  Worschech, A. Forchel, and Y. Yamamoto, Nature Physics {\bf 7},  681  (2011).

\bibitem{coish05}
W.~A. Coish and D. Loss, Phys. Rev. B {\bf 72},  125337  (2005).

\bibitem{johnson05}
A. Johnson, J. Petta, J. Taylor, A. Yacoby, M. Lukin, C. Marcus, M. Hanson, and
  A. Gossard, Nature {\bf 435},  925  (2005).

\bibitem{pfund07}
A. Pfund, I. Shorubalko, K. Ensslin, and R. Leturcq, Phys. Rev. Lett. {\bf 99},
   036801  (2007).

\bibitem{stepanenko12}
D. Stepanenko, M. Rudner, B.~I. Halperin, and D. Loss, Phys. Rev. B {\bf 85},
  075416  (2012).

\bibitem{maune12}
B.~M. Maune, M.~G. Borselli, B. Huang, T.~D. Ladd, P.~W. Deelman, K.~S.
  Holabird, A.~A. Kiselev, I. Alvarado-Rodriguez, R.~S. Ross, A.~E. Schmitz, M.
  Sokolich, C.~A. Watson, M.~F. Gyure, and A.~T. Hunter, Nature {\bf 481},  344
   (2012).

\bibitem{levy02}
J. Levy, Phys. Rev. Lett. {\bf 89},  147902  (2002).

\bibitem{petta05}
J.~R. Petta, A.~C. Johnson, J.~M. Taylor, E.~A. Laird, A. Yacoby, M.~D. Lukin,
  C.~M. Marcus, M.~P. Hanson, and A.~C. Gossard, Science {\bf 309},  2180
  (2005).

\bibitem{taylor07}
J.~M. Taylor, J.~R. Petta, A.~C. Johnson, A. Yacoby, C.~M. Marcus, and M.~D.
  Lukin, Phys. Rev. B {\bf 76},  035315  (2007).

\bibitem{shulman12}
M.~D. Shulman, O.~E. Dial, S.~P. Harvey, H. Bluhm, V. Umansky, and A. Yacoby,
  Science {\bf 336},  202  (2012).

\bibitem{beenakker91}
C. Beenakker and H. van Houten, Solid State Physics {\bf 44},  1  (1991).

\bibitem{barrett06}
S.~D. Barrett and T.~M. Stace, Phys. Rev. B {\bf 73},  075324  (2006).

\bibitem{stace04}
T.~M. Stace, S.~D. Barrett, H.-S. Goan, and G.~J. Milburn, Phys. Rev. B {\bf
  70},  205342  (2004).

\bibitem{nielsen00}
M.~A. Nielsen and I.~L. Chuang, {\em Quantum Computation and Quantum
  Information} (Cambridge University Press, Cambridge, 2000).

\bibitem{averin05a}
D.~V. Averin and E.~V. Sukhorukov, Phys. Rev. Lett. {\bf 95},  126803  (2005).

\bibitem{meunier06}
T. Meunier, I.~T. Vink, L.~H.~W. van Beveren, F.~H.~L. Koppens, H.~P. Tranitz,
  W. Wegscheider, L.~P. Kouwenhoven, and L.~M.~K. Vandersypen, Phys. Rev. B
  {\bf 74},  195303  (2006).

\bibitem{rogge05}
M.~C. Rogge, B. Harke, C. Fricke, F. Hohls, M. Reinwald, W. Wegscheider, and
  R.~J. Haug, Phys. Rev. B {\bf 72},  233402  (2005).

\bibitem{barthel09}
C. Barthel, D.~J. Reilly, C.~M. Marcus, M.~P. Hanson, and A.~C. Gossard, Phys.
  Rev. Lett. {\bf 103},  160503  (2009).

\bibitem{cassidy07}
M.~C. Cassidy, A.~S. Dzurak, R.~G. Clark, K.~D. Petersson, I. Farrer, D.~A.
  Ritchie, and C.~G. Smith, Appl. Phys. Lett. {\bf 91},  222104  (2007).

\bibitem{bylander05}
J. Bylander, T. Duty, and P. P.Delsing, Nature {\bf 434},  361  (2005).

\bibitem{elzerman04}
J.~M. Elzerman, R. Hanson, L.~H. {Willems van Beveren}, B. Witkamp, L.~M.~K.
  Vandersypen, and L.~P. Kouwenhoven, Nature {\bf 430},  431  (2004).

\bibitem{marcinowski11}
{\L}. Marcinowski, M. Krzy{\.z}osiak, K. Roszak, P. Machnikowski, R. Buczko,
  and J. Mostowski, Acta Phys. Pol. A {\bf 119},  640  (2011).

\bibitem{grodecka08a}
A. Grodecka, P. Machnikowski, and J. F{\"o}rstner, Phys. Rev. B {\bf 78},
  085302  (2008).

\bibitem{reilly07}
D.~J. Reilly, C.~M. Marcus, M.~P. Hanson, and A.~C. Gossard, Applied Physics
  Letters {\bf 91},  162101  (2007).

\bibitem{roszak09}
K. Roszak and P. Machnikowski, Phys. Rev. B {\bf 80},  195315  (2009).

\bibitem{gurvitz96}
S.~A. Gurvitz and Y.~S. Prager, Phys. Rev. B {\bf 53},  15932–15943  (1996).

\bibitem{goan01}
H.-S. Goan and G.~J. Milburn, Phys. Rev. B {\bf 64},  235307  (2001).

\bibitem{krummheuer02}
B. Krummheuer, V.~M. Axt, and T. Kuhn, Phys. Rev. B {\bf 65},  195313  (2002).

\bibitem{vagov02a}
A. Vagov, V.~M. Axt, and T. Kuhn, Phys. Rev. B {\bf 66},  165312  (2002).

\bibitem{vagov03}
A. Vagov, V.~M. Axt, and T. Kuhn, Phys. Rev. B {\bf 67},  115338  (2003).

\bibitem{grodecka05a}
A. Grodecka, L. Jacak, P. Machnikowski, and K. Roszak,  in {\em Quantum Dots:
  Research Developments}, edited by P.~A. Ling (Nova Science, NY, 2005), p.\
  47.

\bibitem{roszak10}
K. Roszak, P. Horodecki, and R. Horodecki, Phys. Rev. A {\bf 81},  042308
  (2010).

\bibitem{mahan00}
G.~D. Mahan, {\em Many-Particle Physics} (Kluwer, New York, 2000).

\bibitem{mahan72}
G.~D. Mahan,  in {\em Polarons in Ionic Crystals and Polar Semiconductors},
  edited by J.~T. Devreese (North-Holland, Amsterdam, 1972).

\bibitem{roszak05b}
K. Roszak, A. Grodecka, P. Machnikowski, and T. Kuhn, Phys. Rev. B {\bf 71},
  195333  (2005).

\bibitem{breuer02}
H.-P. Breuer and F. Petruccione, {\em The Theory of Open Quantum Systems}
  (Oxford University Press, Oxford, 2002).

\bibitem{gustavsson08}
S. Gustavsson, I. Shorubalko, R. Leturcq, S. Schön, and K. Ensslin, Appl.
  Phys. Lett. {\bf 92},  152101  (2008).

\bibitem{meystre07}
P. Meystre and M. Sargent, {\em {Elements of Quantum Optics}} (Springer-Verlag,
  Berlin, 2007).

\bibitem{goan01a}
H.-S. Goan, G.~J. Milburn, H.~M. Wiseman, and H. Bi~Sun, Phys. Rev. B {\bf 63},
   125326  (2001).

\bibitem{wiseman01}
H.~M. Wiseman, D.~W. Utami, H.~B. Sun, G.~J. Milburn, B.~E. Kane, A. Dzurak,
  and R.~G. Clark, Phys. Rev. B {\bf 63},  235308  (2001).

\bibitem{gardiner04}
C. Gardiner and P. Zoller, {\em Quantum Noise} (Springer, Berlin, 2004).

\bibitem{blanter01}
Y.~M. Blanter and M. B{\"u}ttiker, Physics Reports {\bf 336},  1   (2000).

\bibitem{fano47}
U. Fano, Phys. Rev. {\bf 72},  26  (1947).

\bibitem{flindt10}
C. Flindt, T. Novotn{\'y}, A. Braggio, and A.-P. Jauho, Phys. Rev. B {\bf 82},
  155407  (2010).

\bibitem{roszak12}
N. Ubbelohde, K. Roszak, F. Hohls, N. Maire, R.~J. Haug, and T. Novotny, Sci.
  Rep. {\bf 2},  374  (2012).

\end{thebibliography}

\end{document}